\documentclass[prd,twocolumn,showpacs,preprintnumbers,floats,floatfix,nofootinbib]{revtex4}
\usepackage{color}
\usepackage{graphicx}
\usepackage{dcolumn}
\usepackage{txfonts}
\newcommand{\dd}{{\rm d}}
\newcommand{\be}{\begin{equation}}
\newcommand{\en}{\end{equation}}
\newcommand{\bea}{\begin{eqnarray}}
\newcommand{\ena}{\end{eqnarray}}
\newcommand{\bean}{\begin{eqnarray*}}
\newcommand{\enan}{\end{eqnarray*}}

\begin{document}

\title{Brane world in Non-Riemannian Geometry}
\author{R. Maier} \email{rodmaier@cbpf.br}
\author{F. T. Falciano} \email{ftovar@cbpf.br}
\hspace{0.5cm}

\affiliation{
CBPF -- Centro Brasileiro de Pesquisas F\'{\i}sicas -- ICRA department \\
Dr. Xavier Sigaud st. 150, Urca, 22290-180, Rio de Janeiro, Brazil}

\date{\today}

\begin{abstract}
We carefully investigate the modified Einstein's field equation in a four dimensional (3-brane) arbitrary manifold embedded in a five dimensional Non-Riemannian bulk spacetime with a noncompact extra dimension. In this context the Israel-Darmois matching conditions are extended assuming that the torsion in the bulk is continuous. The discontinuity in the torsion first derivatives are related to the matter distribution through the field equation. In addition, we develop a model that describes a flat FLRW model embedded in a $5$-dimensional de Sitter or Anti de Sitter, where a $5$-dimensional cosmological constant emerges from the torsion.
\end{abstract}

\pacs{04.20.Cv, 04.50.-h, 11.10.Kk, 11.25.-w}

\maketitle

\section{Introduction}

For a few decades, brane world models have been an interesting option within extra dimension theories. In this scenario, String Theory is valid at high energies and gravity is defined in a 4+D dimensional manifold. At low energy scales, one expectes to recover conventional gravity  and hence the gravitational field should be mostly confined in the 4 dimensional manifold.

From the perspective of String Theory, brane world are phenomenological models with only 1 extra dimension. Hence, it's assumed that the other dimensions become somehow ignorable and all deviations from low energy physics can be implemented in 5 dimensions.

Perhaps the three most successful models are the Dvali-Gabadadze-Porrati (DGP) and the two types of Randall-Sundrum models (see \cite{langlois}-\cite{brax} for a review in brane world). We shall focus on the Randall-Sundrum RSII type in which is possible to have a noncompact extra dimension if the bulk describes a nonfactorizable geometry \cite{rsii}.

There are different reasons to study brane world models. Beside its String theoretical motivation, brane world have some attractive features and applications. It can, for example, solve the hierarchy problem \cite{lisa} or eliminate some singularity issues \cite{sahni}-\cite{maier1}, even though  still persists the stability problem of the Cauchy horizon in gravitational collapse \cite{chandra,maier}. In cosmology it can lead to inflationary or late time accelerating models \cite{ma}-\cite{brodie}. They are also nice models to study holographic ideas such as the AdS/CFT correspondence which can be implemented in the lowest perturbative order \cite{duff}-\cite{molina}.

In this paper, we propose to include torsion effects in the brane world scenario. Even though, up to date, there is no experimental evidence for introducing torsion in gravity, there are some theoretical arguments in favor to consider torsion fields as a desired component in any spacetime theory \cite{hammond}-\cite{hammond3}. In String Theory, the low energy limit effective Lagrangian has besides the aimed gravitational field, a dilaton and an anti-symmetric field \cite{hammond4, fradkin} in which case the torsion potential can be an anti-symmetric Kalb-Ramond field. Furthermore, if one wants to implemente the local Poincare symmetry as part of a gauge theory then torsion fields are also necessary (see \cite{shapiro}-\cite{hehl} for a review on theories with torsion).

The above mentioned Cauchy problem in gravitational collapse is intrinsicaly related to the affine structure of the manifold. Thus, one might also hope to avoid its divergences while including torsion effects. It has been shown \cite{maier} that brane world corrections to the  Schwarzschild metric tend to attenuate gravitational lensing effects which could be a problem to accord with solar system experimental tests. Therefore, torsion can also play an important role in these matters.

In this first analysis, we shall not be concerned with the origin of the torsion field. We shall consider the torsion as a fundamental tensor that defines the affine structure of the bulk but is otherwise completely independent from the metric tensor. Therefore, for a 5 dimensional bulk, there is no a priori constraint in the 50 components of its torsion field.

The paper is organized as follow. Next section is devoted to define some basic geometrical objects, mainly to fix notation and clarify our convention for the geometrical objects that roughtly follows Wald's book \cite{wald} but with a different metric signature. In section \ref{fields} we derive the Gauss-Codazzi embedding equations assuming 5-d Einstein's equation in the bulk with torsion. Then, in section \ref{junction} we analyze the new junction conditions and relate the extrinsic curvature to the bulk matter distribution and the torsion field. In Section \ref{solution} we construct a specific example by proposing an ansatz for the torsion field where is possible to embbed a FLRW metric in a 5D geometry with constant scalar curvature. Depending on the signature of the extra dimension, the cosmological toy model can describe a static universe or a model with a transition from a deccelaration to acceleration phase. Last section is reserved for comments and final remarks.

\section{Basic Equations}

We shall consider non-riemannian manifolds with torsion and therefore it might be usefull to explicitly define some basic relations inasmuch that the position of the index are now rather important and some of the usual symmetries are lost.

In our convention, any metric eingenvalue associated with a time coordinate can at a point be made $+1$ and with a space coordinate $-1$, i.e. a normalized timelike vector has $\xi^\mu \xi_ \mu=1$. The covariant derivative is defined as
\[
\nabla_{b}\xi^a \equiv \xi^a{}_{,b}+\Gamma^a_{bc}\xi^c \quad ,
\]
and the curvature tensor as
\[
R^a{}_{bcd}\equiv \Gamma^a_{db,c}-\Gamma^a_{cb,d}+\Gamma^a_{cm}\Gamma^m_{db}-\Gamma^a_{dm}\Gamma^m_{cb} \qquad.
\]

Let us consider a $N$-dimensional space endowed with a metric tensor ${^{(N)}g}_{ab}$ and a non-trivial affine structure due to torsion terms ${^{(N)}T}^{a}_{.~bc}$. The connection can be defined as
\be
{^{(N)}\Gamma}^{a}_{~bc}= {^{(N)}\left\{^{\, a}_{bc}\right\}}+{^{(N)}K}^{a}_{.~bc} \quad, \label{eq1}
\en
where ${^{(N)}\left\{^{\, a}_{bc}\right\}}$ is the Christoffel symbol and ${^{(N)}K}^{a}_{.~bc}$ is the contortion tensor. The torsion
\be
{^{(N)}T}^{a}_{.~bc}\equiv {^{(N)}\Gamma}^{a}_{~bc}-{^{(N)}\Gamma}^{a}_{~cb}\label{eq2}
\en
together with the metricity condition  ${^{(N)}\nabla}_{c}{g}_{ab}=0$
allow us to write the contortion as
\be \label{eq3}
{^{(N)}K}_{abc}=\frac{1}{2}({^{(N)}T}_{abc}+{^{(N)}T}_{bac}+{^{(N)}T}_{cab})\quad ,
\en
which has the anti-symmetry ${^{(N)}K}_{abc}=-{^{(N)}K}_{cba}$. The curvature tensor can be separeted in its Riemannian and non-Riemannian parts as
\be\label{eq4}
{^{(N)}R}^{a}_{~bcd}={^{(N)}\tilde{R}}^{a}_{~bcd}+{^{(N)}K}^{a}_{~bcd} \quad ,
\en
with ${^{(N)}\tilde{R}}^{a}_{~bcd}$ being the Riemanian tensor defined only with the Christoffels \cite{wald, eisenhart} and
\bea
{^{(N)}K}^{a}_{~bcd}&=&{^{(N)}D}_{c}{^{(N)}K}^{a}_{.~db}-{^{(N)}D}_{d}{^{(N)}K}^{a}_{.~cb}+ \nonumber \label{eq5} \\
&&\quad  +{^{(N)}K}^{m}_{.~db}{^{(N)}K}^{a}_{.~cm}-{^{(N)}K}^{m}_{.~cb}{^{(N)}K}^{a}_{.~dm}\quad ,
\ena
where ${^{(N)}D}$ means covariant derivative constructed only with the Christoffel symbols.

In our study, we consider a bulk manifold $U_{5}$ with coordinates $\{Y^A,~A=0,..,4\}$ and the brane $V_{4}$ as a subspace of $U_5$ with coordinates $\{x^\alpha,~\alpha=0,..,3\}$. We can define an unitary vector field $X^A \in U_{5}$ orthogonal to $V_{4}$. That is,
\be
\label{eqI.4}
{^{(5)}g}_{AB}Y^A_{,~\alpha}X^B=0 \quad,
\en
where $Y^A_{,~\alpha}\, \in V_4 $ forms a vector basis, and
\be
{^{(5)}g}_{AB}X^AX^B=\epsilon=\pm 1\quad ,\label{eqI.3a}
\en
where $\epsilon=+1$ for a timelike extra dimension and  $\epsilon=-1$ for a spacelike extra dimension. Thus, the induced metric in $V_4$ is defined as
\be
^{(4)}g_{\alpha\beta}={^{(5)}g}_{AB}Y^A_{,~\alpha}Y^B_{,~\beta}\label{eqI.2}
\en

In addition of $V_{4}$ and $U_{5}$ being metric spaces, throughout this paper we will consider that the torsion components do not vanish, in general, in any of these two manifolds.

\section{Field Equations}\label{fields}

To derive the effective gravitational equation in the brane, we start with Einstein's field equation in the bulk without cosmological constant, i.e.
\be
{^{(5)}}G_{AB} \equiv {^{(5)}}\tilde{G}_{AB}+{^{(5)}L}_{AB}=\kappa^2_{5}\, {^{(5)}T}_{AB}\quad ,\label{eq6}
\en
where
\bea
^{(5)}\tilde{G}_{AB}&\equiv &{^{(5)}\tilde{R}}_{AB}-\frac{1}{2}{^{(5)}\tilde{R}}\, {^{(5)}g}_{AB} \quad,\label{eq7} \\
{^{(5)}L}_{AB} &\equiv &{^{(5)}K}_{AB}-\frac{1}{2}{^{(5)}{K}} \, {^{(5)}g}_{AB} \label{eq8} \quad ,
\ena
with ${^{(5)}K}_{AB}\equiv {^{(5)}K}^{C}_{~ACB}$ and ${^{(5)}K}\equiv {^{(5)}g}^{AB}~{^{(5)}K}_{AB}$.

By defining the extrinsic curvature
\be
\Omega_{\alpha\beta}=-{^{(5)}g}_{AB}Y^{A}_{,~\alpha}Y^{C}_{,~\beta}{^{(5)}\nabla}_{C}X^B\quad ,\label{eq9}
\en
where ${^{(5)}\nabla}$ is the covariant derivative built with the connection (\ref{eq1}), it is straightforward to show that
\be
{^{(4)}R}_{\alpha\beta\gamma\delta}={^{(5)}R}_{ABCD}Y^{A}_{,~\alpha}Y^{B}_{,~\beta}Y^{C}_{,~\gamma}Y^{D}_{,~\delta} +\epsilon(\Omega_{\beta\delta}\Omega_{\alpha\gamma}-\Omega_{\beta\gamma}\Omega_{\alpha\delta})\; , \qquad \label{eq10}
\en
and
\be
{^{(4)}\nabla}_{\gamma}\Omega_{\alpha\beta}-{^{(4)}\nabla}_{\beta}\Omega_{\alpha\gamma}={^{(5)}R}_{ABCD}X^A Y^{B}_{~,\alpha}Y^{C}_{~,\gamma} Y^{D}_{~,\beta}-T^{\sigma}_{.~\gamma\beta}\Omega_{\alpha\sigma} \; . \quad \label{eq11}
\en

Equations (\ref{eq10}) and (\ref{eq11}) are the Gauss-Codazzi equations for nonvanishing torsion components. Contracting $\beta$ and $\delta$ in the Gauss equation we get
\bea
{^{(4)}G}_{\alpha\gamma}&=&{^{(5)}R}_{AC}Y^A_{,~\alpha}Y^C_{,~\gamma}-\epsilon \, {^{(5)}R}_{ABCD}Y^A_{,~\alpha} X^B  Y^C_{,~\gamma} X^D +\nonumber\\
&&\qquad \quad +\epsilon(\Omega_{\alpha\gamma} \Omega- \Omega_{\alpha\delta}\Omega^{\delta}_{~\gamma})-\frac{1}{2}{^{(4)}g}_{\alpha\gamma}{^{(4)}R} \qquad \quad \label{eq1n}
\ena
with ${^{(4)}G}_{\alpha\beta}\equiv {^{(4)}\tilde{G}}_{\alpha\beta}+L_{\alpha\beta}$, and with an another contraction
\be
{^{(4)}R}={^{(5)}R}-2\epsilon \, {^{(5)}R}_{AC}X^A X^C +\epsilon(\Omega^2- \Omega_{\gamma\delta}\Omega^{\delta\gamma}) \quad . \qquad \label{eq2n}
\en

Finally, using this result in (\ref{eq1n}) we have
\bea
{^{(4)}G}_{\alpha\gamma}&=&{^{(5)}G}_{AC}Y^A_{,~\alpha}Y^C_{,~\gamma}-\epsilon \,  {^{(5)}R}_{ABCD}Y^A_{,~\alpha} X^B  Y^C_{,~\gamma} X^D +\nonumber \\
&&+\, \epsilon(\Omega_{\alpha\gamma} \Omega - \Omega_{\alpha\delta}\Omega^{\delta}_{~\gamma})-\frac{1}{2}\epsilon  \, {^{(4)}g}_{\alpha\gamma}(\Omega^2- \Omega_{\beta\delta}\Omega^{\delta\beta})+  \nonumber \\
&&+ \, \epsilon  \, {^{(5)}R}_{AC}X^A X^C {^{(4)}g}_{\alpha\gamma}\quad .  \label{eq3n}\qquad
\ena

Since Einstein's equation determine only the trace part of the curvature tensor, it is usefull to decompose it in terms of its traces and the Weyl tensor, the trace-free part,
\bean
{^{(5)}\tilde{R}}_{ABCD}&=&{^{(5)}C}_{ABCD} +\frac{2}{3}\Big[{^{(5)}g}_{A[C}{^{(5)}\tilde{R}}_{D]B}-{^{(5)}g}_{B[C}{^{(5)}\tilde{R}}_{D]A}\Big]\\
&&\qquad -\frac{1}{6}{^{(5)}\tilde{R}}{^{(5)}g}_{A[C}{^{(5)}g}_{D]B}\quad .
\enan
Using this decomposition, equation (\ref{eq3n}) can now be written as
\bea
{^{(4)}G}_{\alpha\gamma}&=&\left(\frac{2}{3}{^{(5)}\tilde{R}}_{AC}+{^{(5)}K}_{AC}\right)Y^A_{,~\alpha}Y^C_{,~\gamma}+\epsilon(\Omega_{\alpha\gamma} \Omega-\Omega_{\alpha\delta}\Omega^{\delta}_{~\gamma})\nonumber \\
&-&\frac{1}{2}\epsilon(\Omega^2- \Omega_{\beta\delta}\Omega^{\delta\beta}){^{(4)}g}_{\alpha\gamma}-\epsilon E_{\alpha\gamma}-\epsilon J_{\alpha\gamma}+ \label{eq5n}\\
&+&\left[\epsilon \left(\frac{2}{3}{^{(5)}\tilde{R}}_{AC}+{^{(5)}K}_{AC}\right)X^{A}X^{C}-\frac{5}{12}{^{(5)}\tilde{R}}-\frac12{^{(5)}K}\right]{^{(4)}g}_{\alpha\gamma}
\;, \nonumber
\ena
where we have defined
\bean
E_{\alpha\gamma}&\equiv& {^{(5)}C}_{ABCD} Y^A_{,~\alpha} X^{B} Y^C_{,~\gamma} X^D \quad ,\\
J_{\alpha\gamma}&\equiv& {^{(5)}K}_{ABCD} Y^A_{,~\alpha} X^B Y^C_{,~\gamma} X^D \quad ,
\enan
The bulk field equation (\ref{eq6}) can be used to substitute the trace part of the curvature tensor by the energy-momentum of the bulk and the torsion terms. Therefore, we can rewrite equation (\ref{eq3n}) as
\bea
{^{(4)}G}_{\alpha\gamma}=\frac{2}{3}\kappa^2_{5}\left[{^{(5)}T}_{AC}Y^A_{,~\alpha}Y^C_{,~\gamma}+\left(\epsilon{^{(5)}T}_{AC}X^A X^C-\frac{1}{4}{^{(5)}T}\right){^{(4)}g}_{\alpha\gamma}\right]+&&\nonumber\\
+\epsilon(\Omega_{\alpha\gamma} \Omega-\Omega_{\alpha\delta}\Omega^{\delta}_{~\gamma})-\frac{1}{2}\epsilon{^{(4)}g}_{\alpha\gamma}(\Omega^2- \Omega_{\beta\delta}\Omega^{\delta\beta})- \epsilon \left(E_{\alpha\gamma}+J_{\alpha\gamma}\right) &&\nonumber\\
+\frac13{^{(5)}K}_{AC}Y^A_{,~\alpha}Y^C_{,~\gamma}+\frac13\left(\epsilon {^{(5)}K}_{AC}X^AX^C-\frac{1}{4}{^{(5)}K} \right){^{(4)}g}_{\alpha\gamma}\,.\quad && \label{eq6n}
\ena

These are the modified Einstein's field equation in the brane when one considers nonvanishing torsion for the bulk. The torsion manifest itself introducing extra correction terms in the field equation but also inducing a torsion tensor in the brane. Recall that there is also torsion terms within ${^{(4)}G}_{\alpha\gamma}$ similarly to equation (\ref{eq6}).

To describe the evolution of the field restricted to the brane we still have to specify how the brane is curved with respect to the bulk, i.e. determine the extrinsic curvature. Therefore, next section is devoted to establish the junction conditions to connect the extrinsic curvature to the matter distribution.

\section{Junction Conditions}\label{junction}

Let us assume a given matter distribution restricted to the $4$-dimensional brane (1+3) embedded in a $5$-dimensional bulk space where the extra dimension
can be timelike or spacelike.

In General Relativity, we know that the Israel junction conditions must be satisfied in order to properly describe the geometry of spacetime taking into account possible discontinuities of the matter distribution \cite{israel}. Since we have a $5$-dimensional Einstein equation that connect matter distribution with geometry, there are also consistency conditions relating the extrinsic curvature with discontinuities of the energy-momentum tensor across the brane.

If one assumes that the metric is continuous in the bulk, any discontinuity of its first derivative across the brane must be perpendicular to the brane
\[
\left[ {^{(5)}g}_{AB,C}\right]_{V_4}=\chi_{AB}X_C \qquad ,
\]
where $\left[ f \right]_{V_4}$ means discontinuity of $f$ across $V_4$ in the Hadamard sense \cite{boillat}-\cite{hadamard}, and $\chi_{AB}=\chi_{BA}$. Working this discontinuity up to the curvature tensor, we will be able to connect it with the matter discontinuity through Einstein's equation. However, we still have to specify how the torsion changes due to a matter discontinuity. The torsion modifies the affine structure of the manifold, hence, it should be considered as fundamental as and independent from the metric. Notwithstanding, the field equation shows that its first derivative should be discontinuous if we consider matter discontinuities. Therefore, it seems reasonable to assume that the torsion or equivalently the contortion tensor is continuous just as the metric tensor but its first derivative is discontinuous.

In order to obtain the junction conditions, we will consider Gauss' equation (\ref{eq10}) for a gaussian coordinate system given by
\begin{eqnarray}
\label{eq7n}
ds^2=\epsilon dy^2+ {^{(4)}g}_{\alpha\beta} (x^{\gamma}) dx^{\alpha} dx^{\beta},
\end{eqnarray}
where $y$ denotes the extra dimension and $X^A\equiv \delta^A_{y}$. Therefore, by contracting $\alpha$ and $\gamma$ in (\ref{eq10}) we get
\be
{^{(4)}R}_{\beta\delta}={^{(5)}R}_{AB} Y^{A}_{,~\beta}Y^{B}_{,~\delta}-\epsilon {^{(5)}R}_{yByD} Y^{B}_{,~\beta}Y^{D}_{,~\delta}+\epsilon(\Omega_{\beta\delta} \Omega-\Omega_{\beta\gamma}\Omega^{\gamma}_{~\delta}) \label{eq8n}\quad . \quad
\en
For this coordinate system the Christoffel's symbol are simply
\bean
{^{(5)}\tilde{\Gamma}}^{y}_{By}=0\; , \quad {^{(5)}\tilde{\Gamma}}^{y}_{BC}=-\frac{\epsilon}{2} {^{(5)}g}_{BC,y}\; , \quad {^{(5)}\tilde{\Gamma}}^{A}_{By}=\frac{1}{2} {^{(5)}g}^{AC}{^{(5)}g}_{BC,y} \; \,, \quad
\enan
which will give for the riemannian part of the curvature tensor
\[
{^{(5)}\tilde{R}}_{yByD}=\frac{1}{4}{^{(5)}g}^{CE}{^{(5)}g}_{CD,y} {^{(5)}g}_{BE,y}-\frac{1}{2} {^{(5)}g}_{BD,y,y}\quad .
\]

On the other hand, we can explicitly calculate the derivative of the extrinsic curvature which gives
\[
\Omega_{\alpha\beta,y}=[{^{(5)}K}_{yDB,y}-\frac{1}{2}{^{(5)}g}_{BD,y,y}]Y^{B}_{~,\alpha}Y^{D}_{~,\beta}\quad .
\]
Using the above two equations, one can easily show that
\bean
&&{^{(5)}{R}}_{yByD}Y^{B}_{~,\alpha}Y^{D}_{~,\beta}\; =\; \Omega_{\alpha\beta,y}-{^{(5)}K}_{yyB,D}Y^{B}_{~,\alpha}Y^{D}_{~,\beta}+\\
&&+\Big[\frac{1}{4}{^{(5)}g}^{CE}{^{(5)}g}_{CD,y} {^{(5)}g}_{BE,y}-\frac{1}{2}{^{(5)}g}^{EF}{^{(5)}g}_{BF,y}K_{yDE} +\tilde{\Gamma}^{E}_{~DB}K_{yyE}+\qquad\\
&&+\frac{1}{2}{^{(5)}g}^{EF}{^{(5)}g}_{DF,y}K_{EyB}+K^{M}_{.~DB}K_{yyM}-K^{M}_{.~yB}K_{yDM}\Big] Y^{B}_{~,\alpha}Y^{D}_{~,\beta} \quad .
\enan
Now equation (\ref{eq8n}) may be rewritten as
\be
{^{(5)}{R}}_{\alpha\beta}=\epsilon\Omega_{\alpha\beta,y}+Z_{\alpha\beta} \quad ,\label{eq14n}
\en
where $Z_{\alpha\beta}$ stands for continuous and bounded terms in a finite region that circumscribe the brane.

As it's commonly done in the brane world scenario, we assume that the $5$-dimensional energy momentum tensor has the form
\be
{^{(5)}{T}}_{AB}={^{(5)}{\tilde{T}}}_{AB}+{^{(5)}{{\cal T}}}_{AB}\delta(y)\quad , \label{eq15n}
\en
where ${^{(5)}{\tilde{T}}}_{AB}$ and ${^{(5)}{{\cal T}}}_{AB}$ denote, respectively, the continuous and discontinuous components of ${^{(5)}{T}}_{AB}$ across the brane. In addition, ${^{(5)}{{\cal T}}}_{AB}$ is assumed to be restricted to the brane, i.e. ${^{(5)}{{\cal T}}}_{AB}X^{A}=0$ and can be decomposed as
\be
{\cal T}_{\alpha\beta}\equiv {^{(5)}{{\cal T}}}_{AB}Y^{A}_{~,\alpha}Y^{B}_{~,\beta}=\tau_{\alpha\beta}-\sigma \; {^{(4)}g}_{\alpha\beta} \quad , \label{eq18nnn}
\en
with $\tau_{\alpha \beta}$ describing the matter content confined in the brane and $\sigma$ is the tension of the brane.

We can use the field equation (\ref{eq6}) to write
\be
{^{(5)}{R}}_{AB}=\kappa^2_{5}\Big[{\cal T}_{AB}-\frac{1}{3}{\cal T}{^{(5)}{g}}_{AB}\Big]\delta(y)+\kappa^2_{5}\Big[\tilde{T}_{AB}-\frac{1}{3}\tilde{T}{^{(5)}{g}}_{AB}\Big]\quad .\label{eq15n2}
\en

Recalling that by hypothesis $Z_{\alpha\beta}$, ${^{(5)}\tilde{T}}_{AB}$ and ${^{(4)}g}_{\alpha\beta}$ are bounded functions around $y=0$ and using equations (\ref{eq14n})-(\ref{eq15n2}), we have
\be
\epsilon\lim_{\xi\rightarrow0}\int^{+\xi}_{-\xi}\Omega_{\alpha\beta,y}dy\, =\, \lim_{\xi\rightarrow0}\int^{+\xi}_{-\xi} {^{(5)}{R}}_{\alpha\beta} dy\, =\,
\kappa^2_{5}\Big[{\cal T}_{\alpha\beta}-\frac{1}{3}{\cal T}{^{(5)}{g}}_{\alpha\beta}\Big]\; .
\en

Taking into account the $Z_2$ symmetry, i.e. $\Omega^+_{\alpha \beta}\left(y\right)=-\Omega^-_{\alpha \beta}\left(-y\right)$ and the fact that ${\cal T}_{\alpha \beta}$ is symmetric, the symmetrical part of the extrinsic curvature is given by
\be
\Omega_{(\alpha\beta)}=\frac{1}{2}\epsilon\kappa^2_{5}\Big[{\cal T}_{\alpha\beta}-\frac{1}{3}{\cal T}{^{(5)}{g}}_{\alpha\beta}\Big]\quad . \label{eq18n}
\en
Its anti-symmetrical part can easily be obtained by using definitions (\ref{eq2}) and (\ref{eq9}). Therefore, the extrinsic curvature can be written in terms of the energy-momentum tensor restricted to the brane, the tension of the brane and of the torsion as
\be
\Omega_{\alpha\beta}=\frac{1}{2}\epsilon\kappa^2_{5}\left[\tau_{\alpha \beta}-\frac13 \left(\tau-\sigma \right)g_{\alpha \beta}\right]+\frac{1}{2}T_{ABC}X^A Y^{B}_{,~\alpha}Y^{C}_{,~\beta} \quad .
\en

We can now collect all these terms and include in equation (\ref{eq6n}). Thus, the modified Einstein's equation in the brane becomes
\bea
{^{(4)}G}_{\alpha\beta}+\Lambda_{4}{^{(4)}g}_{\alpha\beta}=8\pi G_{N} \tau_{\alpha\beta}+ \epsilon \kappa^4_{5}\Pi_{\alpha\beta}+\epsilon F_{\alpha\beta}-\epsilon E_{\alpha\beta}-\epsilon J_{\alpha\beta} &&\nonumber\\
+\frac13 {^{(5)}}L_{AB}Y^{A}_{,~\alpha}Y^{B}_{,~\beta}+\frac13 \left( \epsilon{^{(5)}}L_{AB}X^A X^B-\frac14 {^{(5)}}L\right) {^{(4)}g}_{\alpha \beta}\; , \quad && \label{eq23n}
\ena

where we have defined
\bean
\Lambda_{4}&\equiv& \frac{\epsilon}{12}\kappa^4_{5}\, \sigma^2 \quad ,\qquad G_{N}\equiv \kappa_5^4\frac{\epsilon \, \sigma}{48 \pi} \quad, \\
F_{\alpha\beta}&\equiv& \frac{2}{3}\kappa^2_{5}\left[\epsilon {^{(5)}T}_{AB} Y^{A}_{,~\alpha}Y^{B}_{,~\beta}+ \left({^{(5)}T}_{AB}X^A X^B - \frac{1}{4} \epsilon {^{(5)}T}\right){^{(4)}g}_{\alpha\beta} \right]\quad ,\\
\Pi_{\alpha\beta}&\equiv& -\frac{1}{4}\left(\tau_{\alpha\gamma}+\frac{\epsilon}{\kappa_5^2}T_{A\alpha \gamma}X^A\right)
 \left(\tau^{\gamma}{}_\beta+\frac{\epsilon}{\kappa_5^2}T_B{}^{\gamma}{}_\beta X^B\right)+\\
& &+\frac{1}{8}\left(\tau_{\delta \gamma}\tau^{\gamma\delta}+\frac{1}{\kappa_5^4}T_{A\delta \gamma}T_B{}^{\gamma\delta}X^AX^B\right){^{(4)}g}_{\alpha\beta}+\\
&&+\frac{\tau}{12}\left(\tau_{\alpha\beta}+\frac{\epsilon}{\kappa_5^2}T_{A\alpha \beta}X^A\right) -\frac{1}{24}\tau^2 {^{(4)}g}_{\alpha\beta}+\frac{\epsilon \sigma}{6 \kappa_5^2}T_{A\alpha \beta}X^A\quad .
\enan
As we are not assuming a $5$D cosmological constant, it is natural to expect that the $4$D cosmological constant $\Lambda_4$ depends only on the brane tension $\sigma$ and the $5$D Einstein's constant $\kappa^2_5$ (see \cite{maeda}). On the other hand, as long as Newton's constant  $G_{N}$ has to be positive, we have to take positive tension $\sigma$ for a timelike extra dimension or negative for spacelike, i.e. $\sigma=\epsilon \left| \sigma \right|$. Furthermore, the sign of the induced cosmological constant $\Lambda_4$ is also fixed by the nature of the extra dimension. The tensor $F_{\alpha \beta}$ represents the contribution of the $5$-dimensional energy-momentum tensor and $\Pi_{\alpha \beta}$ are correction terms quadratic in $\tau_{\alpha \beta}$ that are no longer symmetric due to presence of the torsion terms $T_{A\alpha \beta}X^A$. One should also note that ${^{(4)}G}_{\alpha\beta}$ in equation (\ref{eq23n}) include torsion terms, recall equations (\ref{eq6})-(\ref{eq8}). Hence, in general, it is also not symmetric.

We have consistently introduced torsion effects in a $5$-dimensional bulk and derived the modified $4$-dimensional Einstein's equation in the context of brane world models. To complete our analysis, we propose a specific example that allow us to construct a cosmological toy model where the brane is described by the FLRW metric.

\section{Embedding FLRW spacetimes in a Non-Riemannian Manifold}\label{solution}

In this section, we shall construct a solution of the field equation (\ref{eq6}) such that it admits the FLRW metric as subspace. The field equation determines how a given matter distribution ${^{(5)}T}_{AB}$ should modify simultaneously the metric and torsion tensors. Notwithstanding, these are non-linear and very involved equations. Therefore, we shall propose an ansatz for the torsion and metric tensors and show that they indeed satisfy the field equation (\ref{eq6}).

Let us consider an ansatz for the torsion in the bulk as
\bea
T_{ABC}=\alpha \; {^{(5)}}g_{A[B}~\varphi_{,\,C]} \quad  &\Rightarrow & \quad K_{ABC}=\alpha \; {^{(5)}}g_{B[A}~\varphi_{,\,C]} \quad \label{eq24n}
\ena
where $\alpha$ is an arbitrary constant and $\varphi$ is a $5$-dimensional scalar field. Furthermore, we shall assume vacuum configuration in the bulk, i.e. we take the metric to describe a $5$-dimensional spacetime with constant scalar curvature, and ${^{(5)}T}_{AB}=0$. In the coordinate system $(u,v,\chi,\vartheta ,\psi)$ the metric can be written as
\be
\dd s^2=H^2_{\Lambda}  v^2 \dd u^2+\frac{\epsilon}{H^2_{\Lambda}v^2}\dd v^2-v^2[\dd \chi^2+\chi^2\dd \Omega^2]\quad ,\label{eq26n}
\en
where $d\Omega=\dd \vartheta^2+sin^2\vartheta \,\dd \psi^2$ is the solid angle and $H^2_{\Lambda}$ is for the time being only an arbitrary constant. Therefore the 5-d Ricci scalar reads
\be
{^{(5)}R}=-20 \, \epsilon \, H^2_{\Lambda}
\label{eq26.1n}
\en
That is, for a timelike extra dimension we have a $5$D de Sitter spacetime. On the contrary, for a spacelike extra dimension we have a $5$D Anti de Sitter spacetime.
\par
Using the above metric and assuming that $\varphi=\varphi(v)$ with
\[
\frac{\dd \varphi}{\dd v}= -\frac{1}{\alpha v}
\]
straightforward but long calculation shows that
\bean
{^{(5)}\tilde{G}}_{AB}&=&6\, \epsilon H^2_{\Lambda} \, {^{(5)}g}_{AB} \\
{^{(5)}L}_{AB} &\equiv &{^{(5)}K}_{AB}-\frac{1}{2}{^{(5)}{K}} \, {^{(5)}g}_{AB}=-6\, \epsilon H^2_{\Lambda} \, {^{(5)}g}_{AB} \quad .
\enan

Therefore, our ansatz satisfies the field equation (\ref{eq6}). Once we have a $5$-dimensional de Sitter or Anti de Sitter solution, one can verify that the flat $4$-dimensional FLRW spacetime
\be
\dd s^2=\dd t^2-a^2(t)[\dd r^2+r^2(\dd \theta^2+sin^2\theta \,\dd \phi^2)] \quad ,\label{eq27n}
\en
can be embedded in the spacetime (\ref{eq26n}) through the following embedding functions
\bean
&&Y^{0}=\frac{1}{H_{\Lambda}}\int \frac{\dd t}{a}\sqrt{1-\frac{\epsilon\dot{a}^2}{H^2_{\Lambda}a^2}}\quad ,\quad Y^{1}=v(t)=-a(t) \qquad \label{eq28n}\\
&&Y^{2}=\chi~=r\quad ,\quad Y^{3}=\vartheta=\theta\quad ,\quad Y^{4}=\psi=\phi \quad .
\enan

One can also calculate the normal vectors that are given by
\[
X^{A}=\pm\left(\frac{\epsilon \dot{a}}{H^2_{\Lambda}a^2}\,,\, \sqrt{H^2_{\Lambda}a^2-\epsilon\dot{a}^2}\,,\, 0\, ,\, 0\, ,\, 0\right)\quad .
\]

We shall consider that the matter content restricted to the brane is described by a perfect fluid
\be
\tau_{\alpha\beta}=(\rho+p)V_{\alpha}V_{\beta}-p{^{(4)}g}_{\alpha\beta}\quad . \label{eq29np}
\en

Again, after some laborious calculation, one can show that the corrected field equations in the brane read
\begin{eqnarray}
\left(\frac{\dot{a}}{a}\right)^2&=&\frac{2\pi G_{N}}{3}\rho\Big[1-\frac{\epsilon}{2|\sigma|}\rho\Big] -\frac{\Lambda_4}{12}\quad ,\label{eq29nnn}\\
\frac{\ddot{a}}{a}&=&-\frac{2\pi G_{N}}{3}\left[\rho+3p-\frac{\epsilon}{\left|\sigma\right|}\rho\left(2 \rho+3p\right)\right]-\frac16 \Lambda_4\quad . \quad \label{eq29nnnn}
\end{eqnarray}

For a perfect fluid with equation of state $p=\omega \,\rho$, the Codazzi equation reads
\bea
&&\dot{\rho}+\frac{\dot a}{a}(5\rho+6p+\sigma)=0\quad \Rightarrow \nonumber \\
&&\Rightarrow \quad \rho=\rho_0\left(\frac{a_{0}}{a}\right)^{5+6\omega}-\epsilon \frac{\left|\sigma\right|}{5+6\omega} \; , \quad \label{eq29m}
\ena
where $\rho_{0}$ and $a_0$ can be taken respectively as the value of the energy density and scale factor today. Taking the time derivative of equation (\ref{eq29nnn}) together with the above Codazzi equation, one re-obtains  the dynamical equation (\ref{eq29nnnn}). This is a consistency check that reassure that our hypothesis of the torsion tensor being continuous across the brane is well defined.

Considering $\rho$  as a decreasing function of the scale factor, i.e. $\omega > -5/6$, the quadratic term on equation (\ref{eq29nnn}) could eventually become relevant for a timelike extra dimension, $\epsilon=+1$,  and provides a way to avoid the initial singularity. However, the nature of the extra dimension also fixes $\Lambda_4>0$ and one can show that it is impossible to find bouncing solutions with a timelike extra dimension. In fact, this dynamical system has only one solution that is a static universe with $\rho=\sigma$. Then, equation (\ref{eq29m}) fixes the value of the scale factor. This static solution is stable in the sense that the constraint equation (\ref{eq29nnn}) does not allow the system to move away from the point $\rho=\sigma$. One can also calculate all orders of time derivative of the scale factor and show that they all vanish as should be if the system is constrained to be fix in the static solution $\rho=\sigma$.

In the case of a spacelike extra dimension, $\epsilon=-1$, the dynamics changes completely. All the terms on the right-hand side of equation (\ref{eq29nnn}) are now positive definite. Hence, we again don't have bouncing solution. Equation (\ref{eq29m}) shows that in an expanding universe the energy density approaches a positive constant
\begin{displaymath}
\lim_{a\rightarrow \infty}\rho \rightarrow \frac{\sigma}{5+6 \omega} \quad.
\end{displaymath}

In addition, if $\omega>-2/3$, the universe starts in a decelerating expanding phase with small scale factor and very high density energy and eventually evolves into an accelerating phase that will tend asymptotically to a de Sitter like expansion. Thus, a spacelike extra dimension can reproduce the transition from a decelerating phase with $\rho \propto a^{-3}$ for $\omega=-1/3$ to an accelerating regime with an effective cosmological constant $\rho=\frac13 \sigma$ and $\ddot{a}/a=\frac{4\pi G_N}{9}\sigma$.

\section{Conclusion and Perspectives}\label{conclusion}

In the present work, we have studied the modifications in the brane world scenario due to the presence of torsion in the affine structure of the bulk manifold for an arbitrarily extra dimension, $\epsilon=\pm1$. The Gauss-Codazzi equations were barely modified with the appearance of an extra torsion term in the Codazzi equation, but now the extrinsic curvature is no longer symmetric. Assuming a $5$-dimension Einstein-like field equation in the bulk, we derived the $4$-dimensional Einstein's equation with the extra terms depending on the $5$-dimensional energy-momentum tensor, the extrinsic curvature and torsion terms. Considering nonvanishing torsion in the bulk, the torsion introduces extra correction terms in the field equation but also induces a torsion tensor in the brane.

We have implemented the junction conditions, which connect the extrinsic curvature to the matter distribution, assuming as usual that the metric tensor is everywhere continuous. Furthermore, inasmuch as the torsion is an independent tensor and in a sense as fundamental as the metric tensor, we have considered that the torsion is also continuous but its first derivative, which appears in the field equation are discontinuous. The novelty in the junction conditions is related to the anti-symmetric part of the extrinsic curvature given by the $5$-dimension torsion tensor projected into the brane.

The identification of the Newtonian constant, $G_N$, fixes the sign of the tension of the brane with respect to the extra dimension, $\sigma=\epsilon |\sigma|$. In addition, the cosmological constant in the brane $\Lambda_4$ also is fixed and has the same sign of the extra dimension. In our toy model, the de Sitter (or Anti de Sitter) bulk solution comes from an effective cosmological constant related to torsion terms. However, if one defines the $5$-dimensional field equation including from the beginning a 5-d cosmological constant, then $\Lambda_4$ is no longer fixed an in fact does not need to have the same sign as the extra dimension.

Finally, we developed a toy model where the torsion tensor has only a scalar degree of freedom. We have shown that this ansatz is equivalent to an effective cosmological constant allowing the de Sitter (or Anti de Sitter) like solution in the bulk. For a timelike extra dimension, $\epsilon=+1$, there is only a unique solution that describes a static universe. Contrarily to other static solutions in the literature \cite{barrow}-\cite{parisi}, this solution is stable in the sense that Friedmann's equations do not allow any matter perturbation restricting the scale factor to a fixed value. In the case of a spacelike extra dimension, $\epsilon=-1$, the tension of the brane contributes to the energy density so that the assymptotic solution is an ever expanding de Sitter universe, but without a varying tension in the brane as in \cite{wong}-\cite{dvali}.

\section*{ACKNOWLEDGEMENTS}

We would like to thank CNPq of Brazil for financial support.  We would also like to thank `Pequeno Seminario' of CBPF's Cosmology Group for useful discussions, comments and  suggestions.\\

\end{document}